\documentclass[twocolumn]{aastex62}
\turnoffeditone
\usepackage[english]{babel}
\usepackage{blindtext}
\usepackage{enumitem}
\usepackage{mathtools}
\usepackage[flushleft]{threeparttable}

\received{}
\revised{}
\accepted{}
\submitjournal{}

\shorttitle{Fast Calculation of Lensing Properties}
\shortauthors{Oguri}

\begin{document}

\title{Fast Calculation of Gravitational Lensing Properties of
  Elliptical Navarro-Frenk-White and Hernquist Density Profiles}

\email{masamune.oguri@ipmu.jp}

\author[0000-0003-3484-399X]{Masamune Oguri}
\affiliation{Research Center for the Early Universe, University of Tokyo, Tokyo 113-0033, Japan}
\affiliation{Department of Physics, University of Tokyo, Tokyo 113-0033, Japan}
\affiliation{Kavli Institute for the Physics and Mathematics of the Universe (Kavli IPMU, WPI), University of Tokyo, Chiba 277-8582, Japan}

\begin{abstract}
We present a new approach for fast calculation of gravitational
lensing properties, including the lens potential, deflection angles,
convergence, and shear, of elliptical Navarro-Frenk-White (NFW) and
Hernquist density profiles, by approximating them by superpositions of
elliptical density profiles for which simple analytic expressions of
gravitational lensing properties are available. This model achieves
high fractional accuracy better than $10^{-4}$ in the range of the
radius normalized by the scale radius of $10^{-4}-10^3$. These new
approximations are $\sim 300$ times faster in solving the lens
equation for a point source compared with the traditional approach
resorting to expensive numerical integrations, and are implemented in
{\tt glafic} software.
\end{abstract}

\section{Introduction}

Gravitational lensing provides an important means of studying the
Universe. With its purely gravitational nature and well-known
underlying physics, gravitational lensing can be used to measure the
dark matter distribution in galaxies and clusters of galaxies as well
as to probe cosmological parameters
\citep[see e.g.,][for reviews]{bartelmann01,treu10,kneib11,limousin13,treu16,oguri19,umetsu20}.

In many cases, information on mass distributions and cosmological
parameters is extracted by comparing gravitational lensing observables
with model predictions assuming parametrized models of mass
distributions. One of the most popular models that are used for the
analysis of gravitational lensing is the \citet[][hereafter NFW]{navarro97}
profile that is a model to represent the radial density
profile of dark matter halos in $N$-body simulations. On the other
hand, the stellar mass distribution is often modeled by the
\citet{hernquist90} profile that approximates the so-called de
Vaucouleurs profile for elliptical galaxies.

One of the advantages of the NFW and Hernquist profiles is
that analytic expressions of their gravitational lensing properties
are available under the assumption of the spherical symmetry
\citep{bartelmann96,wright00,keeton01}. However, observed galaxies and
dark matter halos are not spherically symmetric but are more like
elliptical in projection on the sky. Therefore calculations of
gravitational lensing properties for elliptical density profiles are
important for a range of applications of gravitational lensing. For
instance, the elliptical NFW density profile has been used to measure
projected ellipticities of cluster-scale dark matter halos
\citep{evans09,oguri10,oguri12,umetsu18,okabe20} and to confirm an
important prediction of the standard cold dark matter model.

A challenge of the gravitational lensing analysis using elliptical
density profiles lies in the fact that analytic expressions of their
gravitational lensing properties are available only for a very limited
number of mass models. For instance, analytic expressions for
elliptical NFW and Hernquist density profiles are not known, and as a
result one has to resort to numerical integrations, which are
computationally expensive, to derive their gravitational lensing
properties \citep{schramm90,keeton01}. 

There are several possible ways to overcome such difficulty, at least
partly. \citet{golse02} consider the elliptical NFW potential, in
which the ellipticity is introduced in isopotential contours rather than
in isodensity contours \citep[see also][]{meneghetti03}. While this
model enables fully analytic calculations of gravitational lensing
properties, is also used for measuring projected ellipticities of
cluster-scale dark matter halos \citep[e.g.,][]{richard10}, and is
widely implemented in strong lens modeling softwares including
{\tt gravlens} \citep{gravlens,keeton10},
{\tt lenstool} \citep{jullo07}, {\tt glafic} \citep{glafic}, and {\tt
  lenstronomy} \citep{lenstronomy}, it
predicts unphysical density distributions (e.g., dumbbell-like
isodensity contours and negative densities) when the ellipticity is
large. \citet{du20} propose a broken power-law model, which can
resemble the NFW and Hernquist profiles, as an analytically tractable
elliptical lens model \citep[see also][]{barkana98,tessore15}. Methods
using Fourier series are also explored in \citet{schneider91} and
\citet{chae02}. 

In this paper, we propose models that approximate elliptical NFW and
Hernquist density profiles and have analytic expressions for
their gravitational lensing potentials, deflection angles, convergence
and shear. This is done by expressing NFW and Hernquist profiles by
superpositions of models with known analytic expressions.  
The idea to describe a distribution by a superposition of analytically
tractable models itself is not new. For instance, \citet{hogg13}
model two-dimensional images of galaxies by mixtures of Gaussians for
fast pixel rendering and fast convolution with the telescope
point-spread function. \citet{baltz09} propose to compute
gravitational lensing properties of elliptical density models by a
superposition of elliptical potential models with different ellipticities. 
\citet{shajib19} explore a method to compute lensing and kinematic
properties of arbitrary elliptical mass models by a superposition of
elliptical Gaussian density profiles. Our work is similar in spirit by
that of \citet{shajib19} but proposes to use a different basis
function whose analytic expression of gravitational lensing properties
is simple and easy to implement in codes.

\section{Elliptical NFW and Hernquist Density Profiles}

\subsection{General Elliptical Mass Model}

First, we consider a family of spherically symmetric density profiles
with a characteristic scale $r_0$
\begin{equation}
  \rho(\boldsymbol{x})=\rho_0\,f\left(\frac{|\boldsymbol{x}|}{r_0}\right).
\end{equation}
The corresponding convergence is also a circular symmetric
and is written as
\begin{equation}
  \kappa(r)=\frac{\rho_0}{\Sigma_{\rm cr}}\int_{-\infty}^{\infty} dz\,f\left(\frac{\sqrt{r^2+z^2}}{r_0}\right)=\frac{\rho_0r_0}{\Sigma_{\rm cr}}\kappa_{\rm dl}\left(\frac{r}{r_0}\right),
\label{eq:convergence}
\end{equation}
where $\Sigma_{\rm cr}$ is a critical surface mass density.
We consider an elliptical mass model for which isodensity contours of
convergence have an elliptical symmetry. Specifically, the ellipticity
is introduced by replacing $r$ in Equation~(\ref{eq:convergence}) to
$v$ as 
\begin{equation}
  r\rightarrow v=\sqrt{qx^2+\frac{y^2}{q}},
\end{equation}
where $q$ is the axis ratio of isodensity contours.
Defining $b_{\rm norm}=\rho_0r_0/\Sigma_{\rm cr}$, convergence reduces
to 
\begin{equation}
  \kappa(x, y)=b_{\rm norm}\kappa_{\rm dl}\left(\frac{\xi}{r_0'}\right),
\end{equation}
where
\begin{equation}
  \xi=\sqrt{x^2+\frac{y^2}{q^2}},
  \label{eq:def_xi}
\end{equation}
\begin{equation}
  r_0'=\frac{r_0}{\sqrt{q}}.
\end{equation}
Gravitational lensing properties of this elliptical model can be
computed by one-dimensional numerical integrations
\citep{schramm90,keeton01}. 

\subsection{NFW Profile}

The three-dimensional NFW radial density profile is given by
\citep{navarro97}
\begin{equation}
  \rho^{\rm NFW}(r)=\frac{\rho_{\rm s}}{(r/r_{\rm s})(1+r/r_{\rm s})^2},
\end{equation}
where $\rho_{\rm s}$ is the characteristic density and $r_{\rm s}$ is
the scale radius. Choosing $\rho_0=4\rho_s$ and $r_0=r_{\rm s}$, we
obtain \citep{bartelmann96}
\begin{equation}
  \kappa_{\rm dl}^{\rm NFW}(u)=\frac{1}{2(u^2-1)}\left[1-F(u)\right],
  \label{eq:kdl_nfw}
\end{equation}
where $F(u)$ is defined by
\begin{eqnarray}
 F(u)=\begin{cases}
{\displaystyle \frac{1}{\sqrt{1-u^2}}{\rm arctanh}\sqrt{1-u^2}}&{\displaystyle (u<1)},\\
{\displaystyle \frac{1}{\sqrt{u^2-1}}\arctan\sqrt{u^2-1}}&{\displaystyle (u>1)}.
\end{cases}
\label{eq:hern_fu}
\end{eqnarray}

\subsection{Hernquist Profile}

The three-dimensional Hernquist radial density profile is given by
\citep{hernquist90}
\begin{equation}
  \rho^{\rm Hern}(r)=\frac{M}{2\pi(r/r_{\rm b})(1+r/r_{\rm b})^3},
  \label{eq:kdl_hern}
\end{equation}
where $M$ is the total mass and $r_{\rm b}$ is is the scale and is
known to be related with the half-mass radius $R_{\rm e}$ of the
project surface mass density as $r_{\rm b}=0.551R_{\rm e}$. Choosing
$\rho_0=M/(2\pi r_{\rm b}^3)$ and $r_0=r_{\rm b}$, we obtain \citep{keeton01}
\begin{equation}
  \kappa_{\rm dl}^{\rm Hern}(u)=\frac{1}{(u^2-1)^2}\left[-3+(2+u^2)F(u)\right],
\end{equation}
where $F(u)$ is defined in Equation~(\ref{eq:hern_fu}).

\section{Approximating NFW and Hernquist Profiles}

\subsection{Cored Steep Ellipsoid as a Basis Function}
In this paper, we propose to approximate
$\kappa_{\rm dl}^{\rm NFW}(u)$ (equation~\ref{eq:kdl_nfw}) and
$\kappa_{\rm dl}^{\rm Hern}(u)$ (equation~\ref{eq:kdl_hern}) as
\begin{equation}
  \kappa_{\rm dl}^{\rm NFW}(u)\approx\sum_{i=1}^{N_{\rm NFW}}
  A^{\rm NFW}_i\kappa^{\rm CSE}_{\rm dl}(u; S^{\rm NFW}_i),
  \label{eq:approx_nfw}
\end{equation}
\begin{equation}
  \kappa_{\rm dl}^{\rm Hern}(u)\approx\sum_{i=1}^{N_{\rm Hern}}
  A^{\rm Hern}_i\kappa^{\rm CSE}_{\rm dl}(u; S^{\rm Hern}_i),
  \label{eq:approx_hern}
\end{equation}
where $\kappa_{\rm dl}^{\rm CSE}(u)$ has the following functional form
\begin{equation}
  \kappa^{\rm CSE}_{\rm dl}(u; s)=\frac{1}{2(s^2+u^2)^{3/2}},
\end{equation}
which is an unnamed density profile studied by \citet{keeton98} and
\citet{keeton01}. For the sake of convenience, in what follows we call
(the elliptical extension of) this model as a cored steep ellipsoid
(CSE). 

Once the above approximations are derived, approximated gravitational
lensing properties of the elliptical NFW and Hernquist density profiles are
fully analytically tractable because analytic expressions of gravitational
lensing properties of the CSE profile are available, as explicitly
shown in Section~\ref{sec:cse_lensing}.

We point out that the use of the CSE profile has several advantages
over the elliptical Gaussian profile used in \citet{shajib19}. One is
that first and second derivatives of the lens potential, which are
main quantities used for solving the lens equation and fitting
observed strong lens systems, are described by simple algebraic
functions and hence can be computed very efficiently in codes, in
contrast to those for the elliptical Gaussian for which analytic
expressions involve special functions that are computationally
expensive. Second, unlike the elliptical Gaussian case, a simple
analytic expression of the lens potential is also available for the
CSE profile.  Third, since the CSE model is less localized than the
elliptical Gaussian model, it is expected that we need a smaller
number of CSE components to fit the NFW and Hernquist profiles. 

\subsection{Gravitational Lensing Properties of Cored Steep Ellipsoid}
\label{sec:cse_lensing}

Here we give explicit expressions of the lens potential as well as
first and second derivatives of the lens potential for the CSE model,
which are originally derived in \citet{keeton98}. Specifically, we
consider an elliptical density profile of the following form
\begin{equation}
  \kappa^{\rm CSE}(x, y)=\kappa^{\rm CSE}_{\rm dl}(\xi; s)
  =\frac{1}{2(s^2+\xi^2)^{3/2}},
\end{equation}
where $\xi$ is defined in Equation~(\ref{eq:def_xi}).
The corresponding lens potential is given by
\begin{equation}
  \phi^{\rm CSE}(x, y)=\frac{q}{2s}\ln\Psi-\frac{q}{s}\ln\left[(1+q)s\right],
  \label{eq:cse_potential}
\end{equation}
where $\Psi$ denotes
\begin{equation}
  \Psi=(\psi+s)^2+(1-q^2)x^2,
\end{equation}
\begin{equation}
  \psi=\sqrt{q^2(s^2+x^2)+y^2}.
\end{equation}
We note that the second term of the right hand side of
Equation~(\ref{eq:cse_potential}) is constant and hence can be
omitted. First and second derivatives of the lens potential are simply 
given by
\begin{equation}
  \phi_x^{\rm CSE}(x, y)=\frac{qx(\psi+q^2s)}{s\psi\Psi},
\end{equation}
\begin{equation}
  \phi_y^{\rm CSE}(x, y)=\frac{qy(\psi+s)}{s\psi\Psi},
\end{equation}
\begin{equation}
  \phi_{xx}^{\rm CSE}(x, y)=\frac{q}{s\Psi}\left[1+\frac{q^2s(q^2s^2+y^2)}{\psi^3}-\frac{2x^2(\psi+q^2s)^2}{\psi^2\Psi}\right],
\end{equation}
\begin{equation}
  \phi_{yy}^{\rm CSE}(x, y)=\frac{q}{s\Psi}\left[1+\frac{q^2s(s^2+x^2)}{\psi^3}-\frac{2y^2(\psi+s)^2}{\psi^2\Psi}\right],
\end{equation}
\begin{equation}
  \phi_{xy}^{\rm CSE}(x, y)=-\frac{qxy}{s\Psi}\left[\frac{q^2s}{\psi^3}+\frac{2(\psi+q^2s)(\psi+s)}{\psi^2\Psi}\right].
\end{equation}
Given these analytic expressions, we can easily derive approximated
gravitational lensing properties of the elliptical NFW and Hernquist
density profiles. For instance, the first $x$-derivative (deflection
angle) of the lens potential of the elliptical NFW density profile is
approximated as 
\begin{equation}
  \phi_x^{\rm NFW}(x, y)\approx r_0'b_{\rm norm}\sum_{i=1}^{N_{\rm NFW}}
  A^{\rm NFW}_i\phi^{\rm CSE}_x\left(\frac{x}{r_0'}, \frac{y}{r_0'};
  S^{\rm NFW}_i\right),
\end{equation}
and other approximated lensing properties can also be obtained in a
similar manner. 

\begin{deluxetable}{lcc}
\tablecaption{Parameters for the NFW profile.\label{table:nfw}} 
\tablewidth{0pt}
\tablehead{
\colhead{$i$} & \colhead{$A_i^{\rm NFW}$} & \colhead{$S_i^{\rm NFW}$}
}
\startdata
 1 & 1.648988e-18 & 1.082411e-06 \\
 2 & 6.274458e-16 & 8.786566e-06 \\
 3 & 3.646620e-17 & 3.292868e-06 \\
 4 & 3.459206e-15 & 1.860019e-05 \\
 5 & 2.457389e-14 & 3.274231e-05 \\
 6 & 1.059319e-13 & 6.232485e-05 \\
 7 & 4.211597e-13 & 9.256333e-05 \\
 8 & 1.142832e-12 & 1.546762e-04 \\
 9 & 4.391215e-12 & 2.097321e-04 \\
10 & 1.556500e-11 & 3.391140e-04 \\
11 & 6.951271e-11 & 5.178790e-04 \\
12 & 3.147466e-10 & 8.636736e-04 \\
13 & 1.379109e-09 & 1.405152e-03 \\
14 & 3.829778e-09 & 2.193855e-03 \\
15 & 1.384858e-08 & 3.179572e-03 \\
16 & 5.370951e-08 & 4.970987e-03 \\
17 & 1.804384e-07 & 7.631970e-03 \\
18 & 5.788608e-07 & 1.119413e-02 \\
19 & 3.205256e-06 & 1.827267e-02 \\
20 & 1.102422e-05 & 2.945251e-02 \\
21 & 4.093971e-05 & 4.562723e-02 \\
22 & 1.282206e-04 & 6.782509e-02 \\
23 & 4.575541e-04 & 1.596987e-01 \\
24 & 7.995270e-04 & 1.127751e-01 \\
25 & 5.013701e-03 & 2.169469e-01 \\
26 & 1.403508e-02 & 3.423835e-01 \\
27 & 5.230727e-02 & 5.194527e-01 \\
28 & 1.898907e-01 & 8.623185e-01 \\
29 & 3.643448e-01 & 1.382737e+00 \\
30 & 7.203734e-01 & 2.034929e+00 \\
31 & 1.717667e+00 & 3.402979e+00 \\
32 & 2.217566e+00 & 5.594276e+00 \\
33 & 3.187447e+00 & 8.052345e+00 \\
34 & 8.194898e+00 & 1.349045e+01 \\
35 & 1.765210e+01 & 2.603825e+01 \\
36 & 1.974319e+01 & 4.736823e+01 \\
37 & 2.783688e+01 & 6.559320e+01 \\
38 & 4.482311e+01 & 1.087932e+02 \\
39 & 5.598897e+01 & 1.477673e+02 \\
40 & 1.426485e+02 & 2.495341e+02 \\
41 & 2.279833e+02 & 4.305999e+02 \\
42 & 5.401335e+02 & 7.760206e+02 \\
43 & 9.743682e+02 & 2.143057e+03 \\
44 & 1.775124e+03 & 1.935749e+03 \\
\enddata
\end{deluxetable}

\begin{deluxetable}{lcc}
\tablecaption{Parameters for the Hernquist profile.\label{table:hern}} 
\tablewidth{0pt}
\tablehead{
\colhead{$i$} & \colhead{$A_i^{\rm Hern}$} & \colhead{$S_i^{\rm Hern}$}
}
\startdata
 1 & 9.200445e-18 & 1.199110e-06 \\
 2 & 2.184724e-16 & 3.751762e-06 \\
 3 & 3.548079e-15 & 9.927207e-06 \\
 4 & 2.823716e-14 & 2.206076e-05 \\
 5 & 1.091876e-13 & 3.781528e-05 \\
 6 & 6.998697e-13 & 6.659808e-05 \\
 7 & 3.142264e-12 & 1.154366e-04 \\
 8 & 1.457280e-11 & 1.924150e-04 \\
 9 & 4.472783e-11 & 3.040440e-04 \\
10 & 2.042079e-10 & 4.683051e-04 \\
11 & 8.708137e-10 & 7.745084e-04 \\
12 & 2.423649e-09 & 1.175953e-03 \\
13 & 7.353440e-09 & 1.675459e-03 \\
14 & 5.470738e-08 & 2.801948e-03 \\
15 & 2.445878e-07 & 9.712807e-03 \\
16 & 4.541672e-07 & 5.469589e-03 \\
17 & 3.227611e-06 & 1.104654e-02 \\
18 & 1.110690e-05 & 1.893893e-02 \\
19 & 3.725101e-05 & 2.792864e-02 \\
20 & 1.056271e-04 & 4.152834e-02 \\
21 & 6.531501e-04 & 6.640398e-02 \\
22 & 2.121330e-03 & 1.107083e-01 \\
23 & 8.285518e-03 & 1.648028e-01 \\
24 & 4.084190e-02 & 2.839601e-01 \\
25 & 5.760942e-02 & 4.129439e-01 \\
26 & 1.788945e-01 & 8.239115e-01 \\
27 & 2.092774e-01 & 6.031726e-01 \\
28 & 3.697750e-01 & 1.145604e+00 \\
29 & 3.440555e-01 & 1.401895e+00 \\
30 & 5.792737e-01 & 2.512223e+00 \\
31 & 2.325935e-01 & 2.038025e+00 \\
32 & 5.227961e-01 & 4.644014e+00 \\
33 & 3.079968e-01 & 9.301590e+00 \\
34 & 1.633456e-01 & 2.039273e+01 \\
35 & 7.410900e-02 & 4.896534e+01 \\
36 & 3.123329e-02 & 1.252311e+02 \\
37 & 1.292488e-02 & 3.576766e+02 \\
38 & 2.156527e+00 & 2.579464e+04 \\
39 & 1.652553e-02 & 2.944679e+04 \\
40 & 2.314934e-02 & 2.834717e+03 \\
41 & 3.992313e-01 & 5.931328e+04 \\
\enddata
\end{deluxetable}

\section{Result}

\subsection{Derivation of $A_i$ and $S_i$}
\label{sec:derive_as}

We determine $A_i$ and $S_i$ in Equations~(\ref{eq:approx_nfw}) and
(\ref{eq:approx_hern}) so that these equations hold accurately.
To do so, we quantify the goodness of fit by the following likelihood
function 
\begin{equation}
  \mathcal{L}=\exp\left[-\frac{1}{2}\sum_j
    \frac{\left\{\kappa_{\rm dl}^{\rm X}(u_j)-\sum_{i=1}^{N_{\rm X}}
      A^{\rm X}_i\kappa^{\rm CSE}_{\rm dl}(u_j; S^{\rm X}_i)\right\}^2}
         {\kappa_{\rm dl}^{\rm X}(u_j)^2\sigma^2}\right],
\end{equation}
where ${\rm X}$ denotes either ${\rm NFW}$ or ${\rm Hern}$, and we set
$\sigma=10^{-4}$ that is an approximate target accuracy. To fit
$\kappa_{\rm dl}(u)$ for a wide range of $u$, $u_j$ runs in the range
$10^{-6}\leq u\leq 10^{3}$ and are equally spaced in logarithmic scale
with an interval of $\Delta(\log_{10}u)=0.05$. For a given
$N_{\rm X}$, we set the initial condition of $S^{\rm X}_i$ such that
they are also equally spaced in logarithmic scale in the same $u$
range, and search the maximum likelihood combining the
Metropolis–Hastings algorithm with several different random seeds and
the downhill simplex method. We repeat this procedure as a function of
$N_{\rm X}$ to find an optimal $N_{\rm X}$ that balances the number
with the accuracy. 

Tables~\ref{table:nfw} and \ref{table:hern} shows the parameters
$A^{\rm X}_i$ and $S^{\rm X}_i$ derived using the method mentioned
above, for the NFW and Hernquist profiles, respectively.
The numbers of the CSE components are $N_{\rm NFW}=44$ and
$N_{\rm Hern}=41$. 
 
\begin{figure}
\begin{center}
 \includegraphics[width=1.0\hsize]{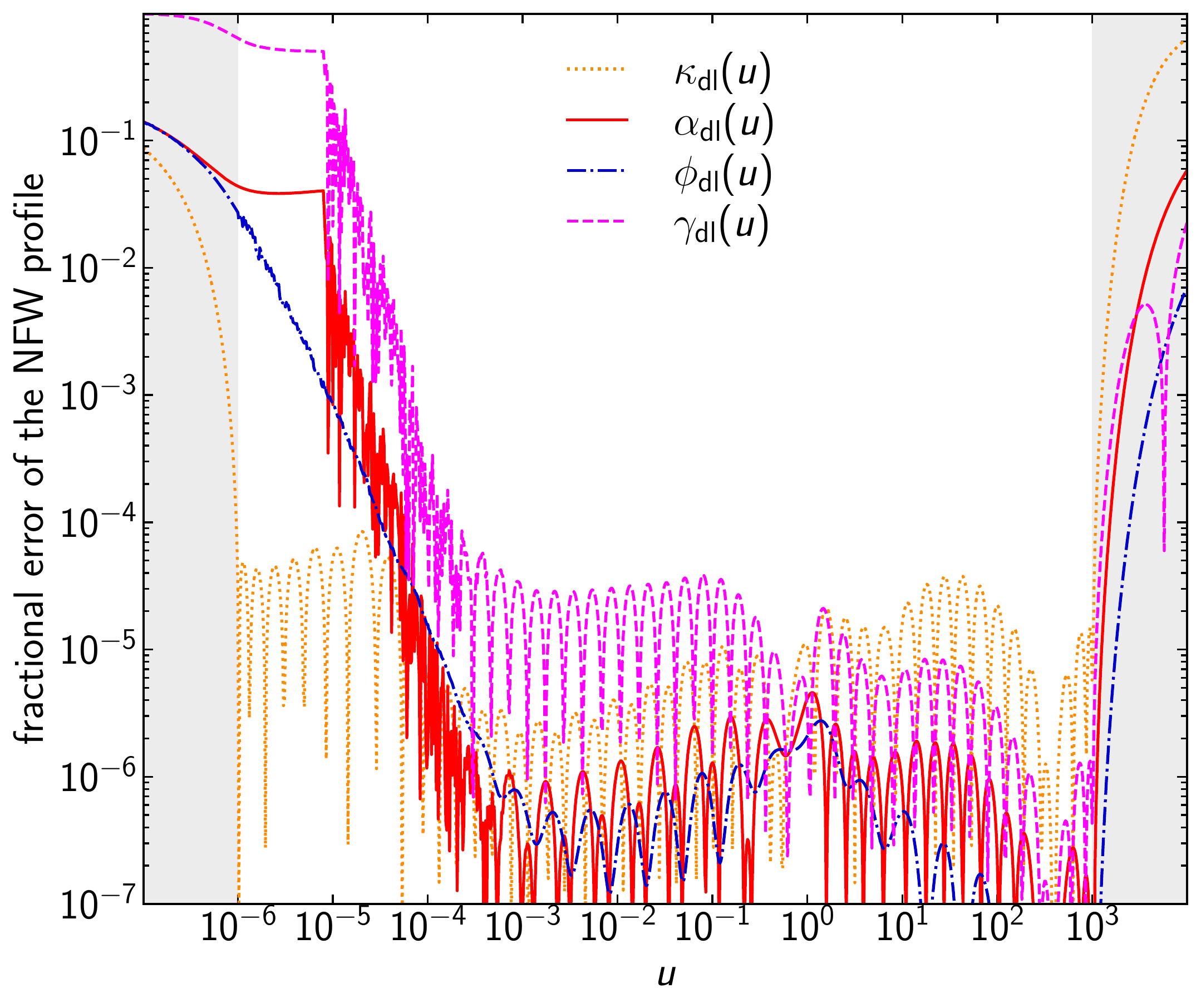}
\end{center}
\caption{The fractional error of the approximation of
  convergence $\kappa_{\rm dl}(u)$ of the NFW profile
  given by Equation~(\ref{eq:approx_nfw}) is shown by
  a dotted line. The parameters $A^{\rm NFW}_i$ and
  $S^{\rm NFW}_i$ are summarized in Table~\ref{table:nfw}. The
  fractional errors of the approximation of the deflection angle
  $\alpha_{\rm dl}(u)$ ({\it solid}), the lens potential
  $\phi_{\rm dl}(u)$ ({\it dot-dashed}), and shear
  $\gamma_{\rm dl}(u)$ ({\it dashed}) are also shown.
  The shaded regions indicate the ranges of $u$ that are not
  used for deriving the parameters $A^{\rm NFW}_i$ and $S^{\rm NFW}_i$.
\label{fig:check_nfw}}
\end{figure}

\begin{figure}
\begin{center}
 \includegraphics[width=1.0\hsize]{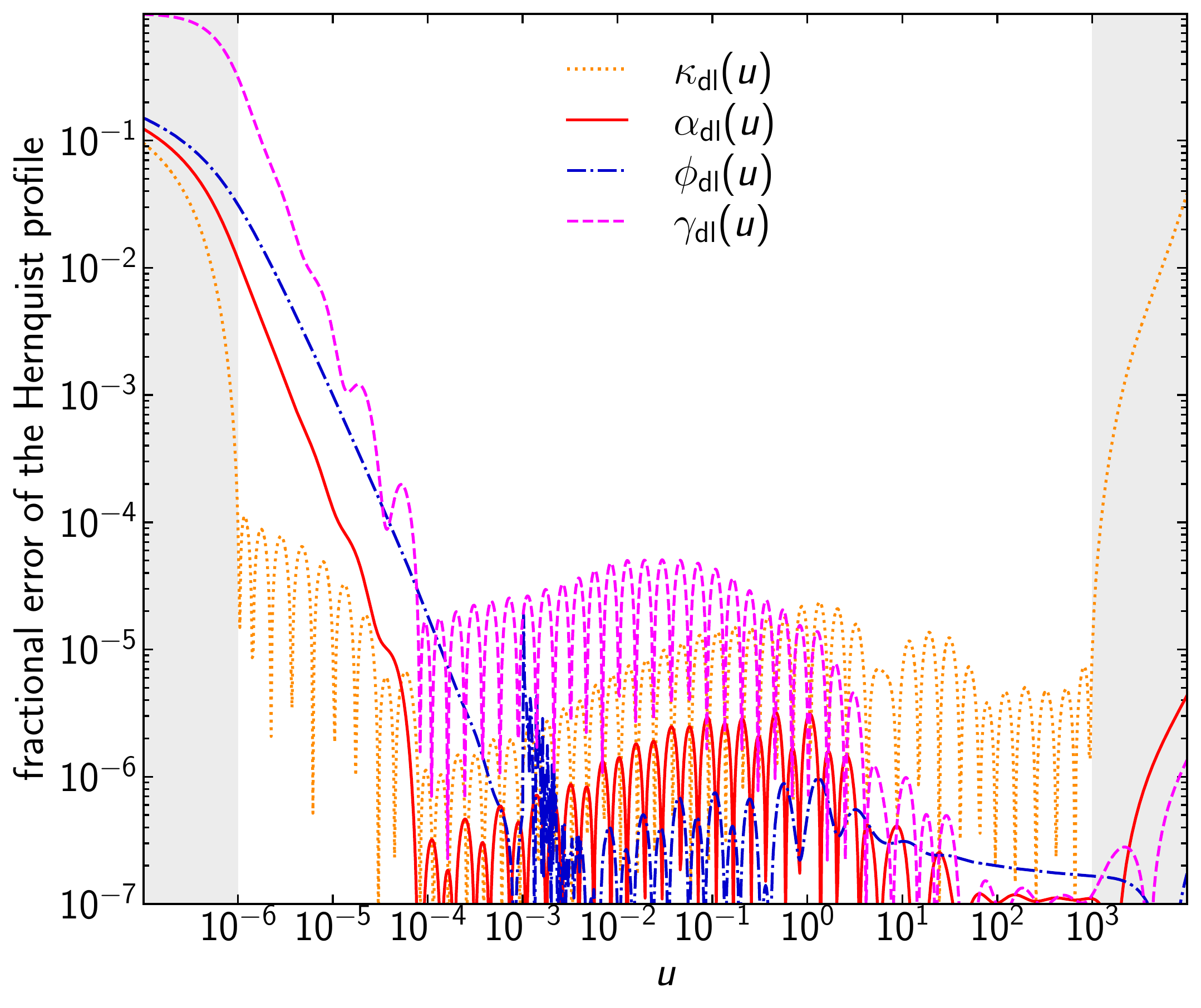}
\end{center}
\caption{Same as Figure~\ref{fig:check_nfw}, but for the Hernquist
  profile.
\label{fig:check_hern}}
\end{figure}

\subsection{Accuracy}

We check the accuracy of our approximations given by
Equations~(\ref{eq:approx_nfw}) and (\ref{eq:approx_hern}) in
Figures~\ref{fig:check_nfw} and \ref{fig:check_hern}, respectively.
In the Figures, we check the accuracy of not only convergence
$\kappa_{\rm dl}(u)$ but also the deflection angle
\begin{equation}
  \alpha_{\rm dl}(u)=\frac{2}{u}\int_0^u u'\kappa_{\rm dl}(u')du',
\end{equation}
the lens potential
\begin{equation}
  \phi_{\rm dl}(u)=2\int_0^u u'\kappa_{\rm dl}(u')
  \ln\left(\frac{u}{u'}\right)du',
\end{equation}
and shear
\begin{equation}
  \gamma_{\rm dl}(u)=\frac{\alpha_{\rm dl}(u)}{u}-\kappa_{\rm dl}(u),
\end{equation}
as all these are important quantities that are frequently used for the
gravitational lensing analysis. We find that our approximation of
convergence is accurate at better than $10^{-4}$ for almost all the
range used for fitting, $10^{-6}<u<10^3$. On the other hand, due to
its non-local nature, the large deviation of convergence at
$u<10^{-6}$ propagates into the deflection angle, the lens
potential, and shear up to $u\sim 10^{-4}$. Hence we conclude that
our approximation using the CSE is accurate at better than $10^{-4}$
for all these lensing properties in the range of the radius normalized
by the scale radius of $10^{-4}-10^3$, both for the NFW and Hernquist
profiles. 

In some cases, accuracy requirements are less strict due to e.g.,
large measurement uncertainties in observations. Our procedure given
in Section~\ref{sec:derive_as} in fact allows us to construct models
with the lower accuracy using the smaller number the CSE components.
As a specific example, we find that it is possible to construct models 
that achieve the accuracy better than $10^{-2}$ in the same range of
the radius normalized by the scale radius of $10^{-4}-10^3$ using the
$N_{\rm NFW}=16$ and $N_{\rm Hern}=13$ CSE components. The parameters
$A^{\rm X}_i$ and $S^{\rm X}_i$ for this lower accuracy case are shown
in Tables~\ref{table:nfw_low} and \ref{table:hern_low}.

\begin{deluxetable}{lcc}
\tablecaption{Parameters for the NFW profile for the lower accuracy
  case.\label{table:nfw_low}} 
\tablewidth{0pt}
\tablehead{
\colhead{$i$} & \colhead{$A_i^{\rm NFW}$} & \colhead{$S_i^{\rm NFW}$}
}
\startdata
 1 & 1.434960e-16 & 4.041628e-06 \\
 2 & 5.232413e-14 & 3.086267e-05 \\
 3 & 2.666660e-12 & 1.298542e-04 \\
 4 & 7.961761e-11 & 4.131977e-04 \\
 5 & 2.306895e-09 & 1.271373e-03 \\
 6 & 6.742968e-08 & 3.912641e-03 \\
 7 & 1.991691e-06 & 1.208331e-02 \\
 8 & 5.904388e-05 & 3.740521e-02 \\
 9 & 1.693069e-03 & 1.153247e-01 \\
10 & 4.039850e-02 & 3.472038e-01 \\
11 & 5.665072e-01 & 1.017550e+00 \\
12 & 3.683242e+00 & 3.253031e+00 \\
13 & 1.582481e+01 & 1.190315e+01 \\
14 & 6.340984e+01 & 4.627701e+01 \\
15 & 2.576763e+02 & 1.842613e+02 \\
16 & 1.422619e+03 & 8.206569e+02 \\
\enddata
\end{deluxetable}

\begin{deluxetable}{lcc}
\tablecaption{Parameters for the Hernquist profile for the lower accuracy
  case.\label{table:hern_low}} 
\tablewidth{0pt}
\tablehead{
\colhead{$i$} & \colhead{$A_i^{\rm Hern}$} & \colhead{$S_i^{\rm Hern}$}
}
\startdata
 1 & 7.775712e-16 & 4.426947e-06 \\
 2 & 3.279878e-13 & 3.551219e-05 \\
 3 & 2.374931e-11 & 1.639271e-04 \\
 4 & 1.137151e-09 & 6.047024e-04 \\
 5 & 5.314908e-08 & 2.180958e-03 \\
 6 & 2.466940e-06 & 7.843573e-03 \\
 7 & 1.125917e-04 & 2.809420e-02 \\
 8 & 4.700637e-03 & 9.893403e-02 \\
 9 & 1.257748e-01 & 3.246017e-01 \\
10 & 9.744152e-01 & 9.409923e-01 \\
11 & 1.434502e+00 & 2.929948e+00 \\
12 & 5.548243e-01 & 1.545137e+01 \\
13 & 6.123431e-01 & 1.883671e+03 \\
\enddata
\end{deluxetable}

\begin{figure}
\begin{center}
 \includegraphics[width=1.0\hsize]{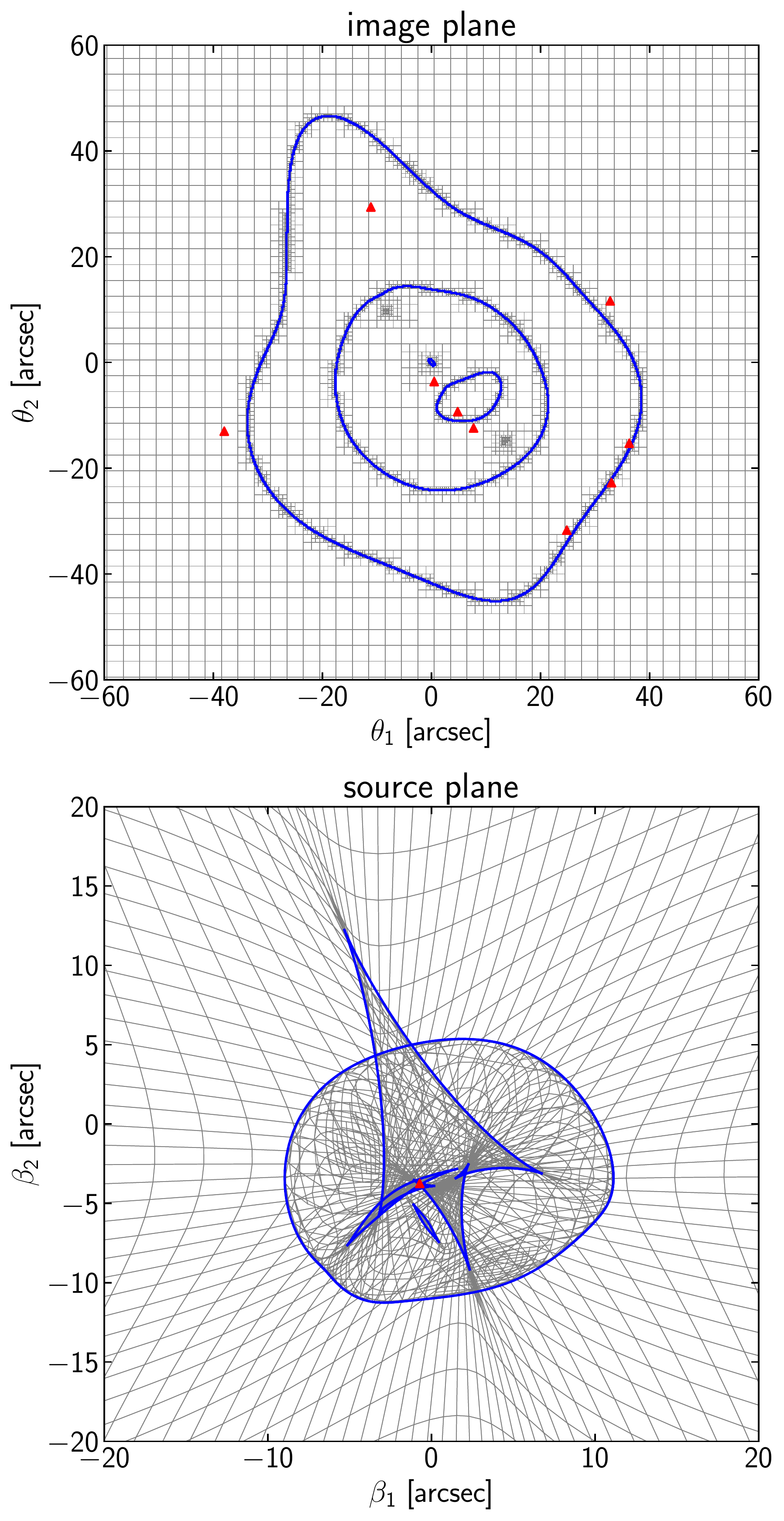}
\end{center}
\caption{Critical curves and caustics of an example of the multiple
  lens plane mass model that is used to check the performance of the
  approximation are shown by solid lines. In this example three
  elliptical NFW and three elliptical Hernquist lenses placed
  over three lens planes at $z=0.3$, $0.6$, and $0.9$ are included in
  total. Filled triangles show the position of a point source at $z=3$
  in the source plane as well as positions of the corresponding multiple
  images in the image plane. Adaptive mesh that is prepared in solving
  the lens equation with {\tt glafic} software \citep{glafic} is also
  shown by gray lines. 
\label{fig:plot_point} }
\end{figure}

\subsection{Demonstration with {\tt glafic}}

The new approximations (the higher accuracy version with
$N_{\rm NFW}=44$ and $N_{\rm Hern}=41$) of the elliptical NFW and
Hernquist density profiles proposed in this paper are implemented in
{\tt glafic} software\footnote{The binary files are available at
\url{https://www.slac.stanford.edu/~oguri/glafic/}, and 
the source code is available at
\url{https://github.com/oguri/glafic2}.}
\citep{glafic} as lens model names of {\tt anfw} and {\tt ahern},
respectively.  The latest version of {\tt glafic} also supports
multiple lens plane gravitational lensing
\citep[e.g.,][]{schneider92}. Adopting three lens plane gravitational
lensing as an example, we explicitly check how our new approximations
speed up gravitational lensing calculations. 

In our specific example, we consider three lens planes at redshift
$z=0.3$, $0.6$, and $0.9$. At each lens plane, we place a compound lens
comprising the elliptical NFW and Hernquist profiles with masses of
$3\times 10^{14}h^{-1}M_\odot$ and $3\times 10^{12}h^{-1}M_\odot$,
respectively. All these NFW and Hernquesit components have non-zero
ellipticties $1-q$ ranging from $0.2$ to $0.8$. 
We derive multiple images of a point source placed at redshift $z=3$
with {\tt glafic}. When solving the lens equation, the initial grid is
set with a spacing of $3''$ ({\tt pix\_poi}\,$=3.0$) and the grid size
is adaptively refined up to fifth level ({\tt maxlev}\,$=5$).  

Figure~\ref{fig:plot_point} shows critical curves, caustics, and
positions of multiple images in this example. We find that shapes
of the critical curve and caustic are almost indistinguishable between
results with and without the approximation. In this specific example,
there are 9 multiple images, and their positions and magnifications
with and without the approximation match typically within $0.02''$ and
$0.5\%$, respectively.  These differences are sufficiently
small compared with the typical accuracy of cluster mass modeling and
hence are unimportant. We note that the differences in fact largely
come from the imperfect accuracy of numerical integrations of the
elliptical NFW and Hernquist profiles without the approximations, as
those numerical integrations are slow to converge and those in
{\tt glafic} are accurate only up to fractional errors of
$\sim 10^{-3}-10^{-4}$ level so as not to be computationally intensive. 

More importantly, the new approximations make these calculations
much faster. With a single core calculation with 2.3 GHz Intel Core
i7, the new approximations allow us to find those multiple images in
0.13~seconds, while the calculations resorting to numerical integrations
take 36~seconds to find those multiple images. Therefore, the new
approximations are $\sim 300$ times faster than the traditional
calculation involving numerical integrations, at least for the case of
this specific example to solve the lens equation.

\section{Summary}

We have presented a new approach for fast calculation of gravitational
lensing properties of the elliptical NFW and Hernquist density
profiles. In this approach, the elliptical NFW and Hernquist density
profiles are approximated by superpositions of the cored steep
ellipsoid for which simple analytic expressions of gravitational
lensing properties are available. These new approximations enable us
to compute gravitational lensing potential, deflection angles,
convergence, and shear at fractional accuracy better than $10^{-4}$ in
the range of the radius normalized by the scale radius of
$10^{-4}-10^3$, which is sufficiently wide for most of applications in
both strong and weak 
gravitational lensing. These new approximations are also $\sim 300$
times faster in solving the lens equation for a point source compared
with the traditional approach involving numerical integrations.
These approximated models are already implemented in {\tt glafic}
software \citep{glafic} and are available for immediate applications.

We note that our approach enables fast calculation of lens potentials
for the elliptical NFW and Hernquist density profiles, which may
become important when we study the wave optics effect for which the
Kirchhoff diffraction integral involving the lens potential needs to
be evaluated \cite[see e.g.,][for a review]{oguri19}. 

\section*{Acknowledgments}

We thank the anonymous referee for useful comments.
This work was supported in part by World Premier International
Research Center Initiative (WPI Initiative), MEXT, Japan, and JSPS
KAKENHI Grant Number JP20H04725, JP20H00181, JP20H05856, JP18K03693.

\bibliography{ref} 

\end{document}